# Mosques Smart Domes System using Machine Learning Algorithms

Mohammad Awis Al Lababede[1]
Computer Science Department
Mutah University
Al Karak, Jordan

Anas H. Blasi[2]
Computer Information Systems Department
Mutah University
Al Karak, Jordan

Mohammed A. Alsuwaiket[3]
Computer Science and Engineering Technology Department
Hafar Batin University
Hafar Batin, Saudi Arabia

*Abstract*—Millions of mosques around the world are suffering some problems such as ventilation and difficulty getting rid of bacteria, especially in rush hours where congestion in mosques leads to air pollution and spread of bacteria, in addition to unpleasant odors and to a state of discomfort during the pray times, where in most mosques there are no enough windows to ventilate the mosque well. This paper aims to solve these problems by building a model of smart mosques' domes using weather features and outside temperatures. Machine learning algorithms such as k-Nearest Neighbors (k-NN) and Decision Tree (DT) were applied to predict the state of the domes (open or close). The experiments of this paper were applied on Prophet's mosque in Saudi Arabia, which basically contains twenty-seven manually moving domes. Both machine learning algorithms were tested and evaluated using different evaluation methods. After comparing the results for both algorithms, DT algorithm was achieved higher accuracy 98% comparing with 95% accuracy for k-NN algorithm. Finally, the results of this study were promising and will be helpful for all mosques to use our proposed model for controlling domes automatically.

## I. Introduction

Islam is the second largest religion after Christianity in the world, according to a study conducted in 2015, Islam have 1.9 million followers in the world, representing 24.8% of the world's population [1]. In addition, there are 3.6 million mosques around the world [2]. In fact, mosques have a major problem of a good ventilation due to the crowd since there are many worshipers inside the mosque, and as the windows are not enough for fully ventilation, in addition to the problem of the presence of bacteria and moisture on the mosques' carpets.

In general, the mosques need to keep up with technology and evolution, even if moving domes are exist in some mosques, they are inefficient and there are many problems that prevent them to stay open for a long time, such as weather change and weather conditions, where the weather suddenly turns from clear to dusty, rain, hurricane problems, sandstorms and strong sunlight that disturb worshipers, also the temperatures are high or low outside, so it is difficult to control these domes manually and there are many factors that make the decision hard to open or close the domes.

This paper introduces solutions to previous problems by creating smart moving domes considering weather prediction. However, weather can be predicted by studying satellite data or the behavior of animals affected by weather changes and weather maps [3]. To make this research applicable and reliable, the experiments in this paper will be applied on Prophet's Mosque in Saudi Arabia, which basically contains twenty-seven manually moving domes. Where a Prophet's Mosque is one of the largest mosques in the world and the second holiest Islamic site which was built by the messenger of Allah; Mohammad -peace be upon him- in 1 Hijra, and then was expanded several times throughout the history by the princes of Islamic countries for each period and the largest expansion was during the reign of Saudi Arabia in 1994. The domes were built using silver, granite, gold and marble. Moreover, Prophet's Mosque is expanded to suit 707.000 worshipers, and it is visited by more than 278,000 Muslim every hour from all over the world [4].

Some requirements have been proposed for the domes system to make them move and some required materials should be considered such as:

- Steel rails: should be made to handle the friction caused by the moving domes, where the domes are placed on these rails for opening and closing.

- Arduino controller: a simple Arduino device is needed to send (on, off) signals, which gets information from a processing device that forecasts weather and dome state.

- Processing device: a computer or processing cloud used to implement the weather prediction algorithm and the dome state.

- Machine learning algorithms: to predict the weather and the domes state, a model should be built using machine learning algorithms such as Decision Tree (DT) and k-Nearest Neighbors (k-NN), where weather factors such as visibility, temperature, humidity, strong wind, clock and barometer will be used to make the decision to open or close the domes.

- Rainfall sensor: to minimize the error rate of the proposed model, it will be used to detect rainfall at the real time and force the domes to close.

In this paper, a model of smart domes will be built using Machine Learning algorithms such as Decision Tree (DT) and k-Nearest Neighbors (k-NN), considering some weather factors like minimum temperature, maximum temperature, wind speed direction, humidity, and average dew point [5]. The main aim of this paper is to solve the problem of ventilation of the mosques by controlling the domes automatically using ML algorithms instead of controlling them manually.

The paper is organized as follows: Section II reviews the related work of weather forecasting using different ML techniques, section III describes the process followed to prepare the data including data understanding, selecting, transforming, and model building. Section IV describes the interpretation and evaluation of the results. Finally, section V discusses the conclusions and draws the future work.

## II. Related Work

In this section, some related work will be presented and reviewed to show how others have applied machine learning techniques such as Decision tree (DT), k-Nearest Neighbors (k-NN), Artificial Neural Networks (ANN), Linear Regression (LR), Naive Bayes (NB) and other data mining techniques [5] to predict the weather changes.

According to the studies [6, 7] Artificial Neural Network (ANN) and Deep learning are gaining much popularity due to its supremacy in terms of accuracy when trained with huge amount of data. Deep learning outshines several other artificial intelligence techniques when there is lack of domain expertise in feature engineering, or when it comes to complex problems such as optimization, image classification, natural language processing, and speech recognition.

Some other studies concern about nature events such as [8], authors studied the micro seismic events were detected and classified through combining ML algorithms such as back up vector machine, MLP, NN, C4.5 decision tree and k-NN in the form of boost learning. The procedure of experiment was in a way that less important and important seismic events caused by weight falling from various heights and different distances of far, middle and near recorded by laboratory devices and sensors. Then, the data were pre-processed and classified. The classification was based on the level of height, distance and considered sensors. The precision and accuracy were significantly improved by this strategy. After simulation of their proposed method, it was observed that the precision of proposed boost method was improved up to 6.1% compare to the other methods. The error rate improved up to 0.82% and the recalling and accuracy of detection and classification to the best answer were also improved in the proposed method up to 2.31% and 6.34%, respectively.

According to [9], the authors have built a hybrid system between a multilayer perceptron (MLP) and Radial basis function (RBF) to enhance weather forecasting in Saudi Arabia by training both the individual neural network and the hybrid network using weather elements that exist in the dataset. The outcome was either rainy or dry, the inputs were appointed to determine correlation coefficient, Root Mean Square Error (RMSE) and scatter index. The paper showed that the hybrid model was better of the individual front grille model (MLP and RBF) and the results were more accurate and had a better learning ability.

In other paper [10], the authors studied a high-precision temperature prediction through complex data for atmospheric. There were two types of weather prediction: dynamic and experimental. They used the Back Propagation Neural Network (BPN) approach and Feed Forward Neural Network and they used randomly weights for all nods to train a data collection using three hidden layers to numerical weather prediction.

According to paper [11], the authors studied forecasting rainfall in India using artificial intelligence techniques to support agriculture and crop multiplication through predicting the precipitation for the next year, they studied a precipitation of historical data and its relationship to the atmosphere using Multiple Linear Regression (MLR) approach. Moreover, the data used is for 30 years from 1973 to 2002 and included data on cloud cover, average temperature and precipitation to Udaipur city, Rajasthan, India. The system was predicting the monthly rainfall quantities, which was very closed to the actual results.

Authors in paper [3] were predicting the weather using arbitrary decision tree algorithm. The elements in the dataset were divided using divide and conquer technique. The data used in this study were obtained from [12] for the city of London. In this study, the weather was predicted by studying satellite data or by the behavior of animals affected by weather changes, weather maps. Finally, using split evaluator, information gain and entropy, they got higher resolution and a small decision tree.

The authors studied in paper [13] a data classification tools and made a detailed comparison between the three tools Decision tree, KNN, and Naive Bayes. They explained the advantages and disadvantages of each one and explained how they are working. Moreover, some examples were given for each type of tools, and applied some examples of weather forecasting. Finally, they concluded that the decision tree was the best and most accurate.

According to the study in [14], authors have tested the applicability of soft computing technique-based rainfall-runoff models (ENN and ANN) to simulate runoff in Bihar. Runoff and antecedent runoff, precipitation, antecedent precipitation over the basin, at three gauging stations in the basin were first identified as appropriate input variables, and then CCF curves at differ time lags were plotted to select the potential input variables. Monthly rainfall data of two stations and discharge data of one station for the period 1986-2014 were utilized as data sets for the development of proposed models. Based on their statistical, it was indices it had been established that ENN outperformed ANN and is more accurate as compared to the traditional ANN method for rainfall-runoff modelling. The results of their study were helpful in selecting the appropriate model for the discharge simulation in Bihar and thereby helping planners for effective flood mitigation.

It can be concluded that there are many researchers have done work related to the weather prediction using machine and deep learning algorithms, but very limited applied their studies for domes specially for mosques. Next section, the methodology of our study will be presented in detail.

### III. METHODOLOGY

In this paper, Knowledge Database Discovery (KDD) methodology has used to present all the steps required to build up the model from the data collection stage to the preprocessing, cleaning, selecting the data, then choosing the appropriate model and finally evaluating the results. The steps of KDD are mentioned in Figure 1.

#### A. Selection

In this paper, Data were collected from Kaggle website [12] for the weather in Saudi Arabia, the dataset contains the hourly changing weather from 2017 to 2019 for all the cities in Saudi Arabia. The size of dataset is 249024 records. However, the selected attributes are date, hour, minute, day, temperature, humidity, wind strength, barometer and visibility.

#### B. Preprocessing

Using the correlation function, the relationship between features has been showed, then some uncorrelated features such as date, time, year, month, day and minute were ignored for the dataset. Al Madina city was chosen as target city. Finally, after cleaning the dataset, 19964 records were left for further processing.

#### C. Transformtion

In this stage, some other columns such as "state" and "New weather" have been added, and then the state of the weather for thirty six state have converted to (0, 1) and added into "new_weather" column, and the "state" column has the final state for domes where 0 means the dome is close and 1 means the dome is open. The following Table I. shows the transformation of the weather attributes.

Temperature could change the state of domes, so if the final domes state "new weather" is 1 and the temp degree is more than 16° and lower than 27°, then the value of state column will stay 1 (keep domes open). Otherwise, the value of state will be changed to 0 (close domes). Figure 2 shows the impact of temperature factor on domes state.

#### D. Modeling

In this section, the weather status and domes state will be predicted using Decision Tree (DT) and k-Nearest Neighbors (k-NN) algorithms. The proposed model can determine the state of the domes through seven factors. When temperature below 16° or more than 27°, the domes must be closed. Otherwise, the domes state must be 1 or 0. As mentioned in Figure 4, case 1 means that domes are open while the case 0 means that domes are close. After determining the state of the domes, the system will give a signal to the Arduino controller device, where the Arduino controller operates the rails to open and close the domes.

Regarding to the air conditioners, if the domes are open, air conditioners will be turned off, but if the domes are open, then the air conditioners will be turned off. Finally, to solve the problem of unexpected rainfall, rain detection sensor will be used, see Figure 3.

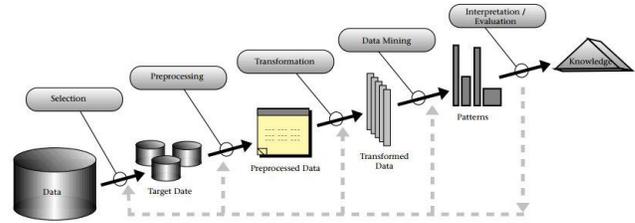

Fig. 1. Knowledge Database Discovery (KDD) Processes [15].

TABLE. I. THE TRANSFORMATION OF THE WEATHER ATTRIBUTES

| State | Visibility | Barometer |
|---|---|---|
| 1 | 1 | Clear |
| 2 | 0 | Sunny |
| 3 | 1 | Passing clouds |
| 4 | 1 | Low level haze |
| 5 | 1 | Scattered clouds |
| 6 | 1 | Partly sunny |
| 7 | 1 | Broken clouds |
| 8 | 0 | Duststorm |
| 9 | 0 | Sandstorm |
| 10 | 1 | Pleasantly warm |
| 11 | 1 | Thunderstorms passing clouds |
| 12 | 1 | Thunderstorms partly sunny |
| 13 | 1 | Thundershowers |
| 14 | 1 | Mostly cloudy |
| 15 | 1 | Thunderstorms Broken clouds |
| 16 | 1 | Thunderstorms Scattered clouds |
| 17 | 0 | Extremely hot |
| 18 | 1 | Mild |
| 19 | 1 | Thunderstorms Partly clouds |
| 20 | 0 | Rain Partly cloudy |
| 21 | 0 | Rain Scattered clouds |
| 22 | 0 | Rain Broken clouds |
| 23 | 1 | Haze |
| 24 | 1 | Overcast |
| 25 | 1 | Dense fog |
| 26 | 0 | Rain passing clouds |
| 27 | 0 | Rain Mostly cloudy |
| 28 | 0 | Rain Partly sunny |
| 29 | 1 | Fog |
| 30 | 0 | Hail Partly sunny |
| 31 | 1 | Thundershowers passing clouds |
| 32 | 1 | More clouds than sun |
| 33 | 1 | Thunderstorms more clouds than sun |
| 34 | 1 | Thunderstorms |
| 35 | 1 | Partly cloudy |
| 36 | 0 | Hail |

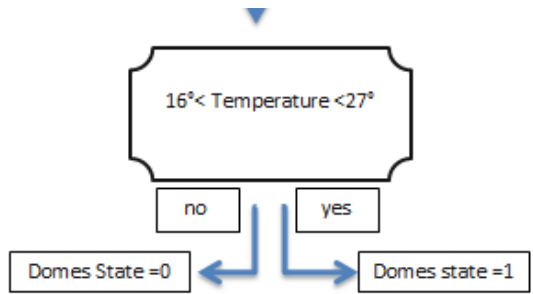

Fig. 2. Impact of Temperature Factor on Domes State.

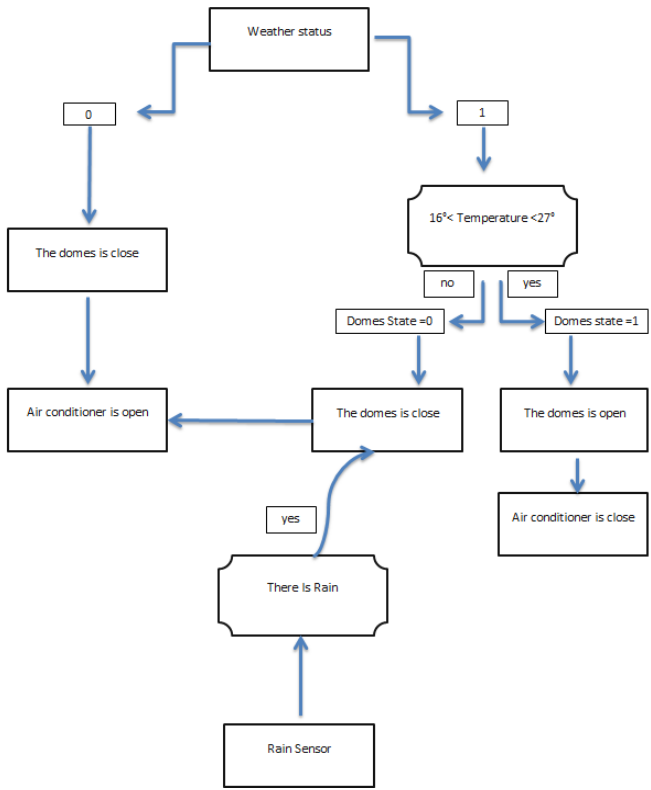

Fig. 3. Proposed Model of Weather and Domes Status.

In this stage, a model will be built and due to the study [13], it has been found that the decision tree and kNN algorithms are most effective than others. Furthermore, both algorithms have high accuracy for weather forecasting, in addition to the speed of training and testing.

Weather forecasting is a complicated process. However, there are two types of weather forecasting techniques, one is the dynamic and the other is experimental. The experimental prediction is used by meteorologists if there are a lot of data and used in a local area, the dynamic prediction is used for broad forecasting and it is not ineffective with short-term.

*1) Decision Tree (DT):* A decision tree is predictive modeling technique used in classification, clustering and prediction tasks. DT is one of the most common machine learning algorithms and it uses divide and conquer technique to split the problem search space into subsets [16].

Python programming language will be used for the implementation stage. However, the prediction model of the weather and domes status will be built using sklearn with some important libraries and methods such as accuracy_score, train_test_split, DecisionTreeClassifier and panads.

The algorithm aims to divide and distribute the records of dataset into depth-first greedy approach or breadth-first approach, where the structure of the algorithm must consist of root, internal nodes and leaf, where each node refers to a condition on the attribute and uses approach from top to bottom. In addition, the decision tree is quick, simple and easy to understand of representation.

After finishing the preprocess of the previous step, data will be entered to the model to be trained to build the model and then the model will be tested. For this step, the target (output) and the attributes (inputs) have been determined. See Table II.

As mentioned in Table 2, the weather features ("Temp", "wind", "humidity", "hour", visibility", "barometer") will be used as inputs to the model, and "state" will be used as output. Data have divided into two splits, 33% for testing and 77% for training. The max leaf nodes are 50 nodes and random state is 324.

After testing the model, the accuracy has been calculated through the accuracy functions which have resulted 98%. In addition, other evaluation methods have been calculated as well. Table III shows the results of evaluation methods for DT.

*2) k-Nearest Neighbors (k-NN):* k-NN algorithm is one of the best algorithms for Machine Learning, which is easy to use, and it is introduces an excellent accuracy compared to other algorithms [17]. k-NN is based on calculating the distance between the point required with all points in the neighborhood, and choosing the shortest distance, which depends on the value of k, where k is the number of neighborhoods must be compared with the required point. K-NN is the fastest technology to learning comparing with neural network, decision tree and Bayes networks, but it takes a long time during the classification process and works well on data with multiple classifications.

In this paper, k-NN is used with 141 k's, where k has calculated by the square root of the data records 19964. Python programming language will be used for the implementation stage. However, the prediction model of the weather and domes status will be built using sklearn with some important libraries and methods such as accuracy_score, train_test_split, KNeighborsClassifier, pandas, matplotlib.

The weather features ("Temp", "wind", "humidity", "hour", visibility", "barometer") have used as inputs to the model, and "state" has used as output. Data has split into 30% for testing and 70% for training, and random state is 101.

After testing the model, the accuracy has been calculated through the accuracy functions which have resulted 95%. In addition, other evaluation methods have been calculated as well. Table IV shows the results of evaluation methods for k-NN.

TABLE. II. SAMPLE OF PREPROCESSED DATA FOR WEATHER FEATURES

| State | Visibility | Barometer | Humidity | Wind | Temp | hour | |
|---|---|---|---|---|---|---|---|
| 1 | 16 | 1020.0 | 0.33 | 0 | 21 | 24 | 0 |
| 1 | 16 | 1020.0 | 0.35 | 9 | 19 | 1 | 1 |
| 1 | 16 | 1020.0 | 0.37 | 11 | 19 | 2 | 2 |
| 1 | 16 | 1020.0 | 0.40 | 7 | 18 | 3 | 3 |
| 1 | 16 | 1019.0 | 0.39 | 0 | 17 | 4 | 4 |

TABLE. III. THE RESULTS OF EVALUATION METHODS FOR DT

| | F1→1 | F1→0 | Weighted Avg → F1 | MSE | Accuracy |
|---|---|---|---|---|---|
| Decision Tree | 0.97 | 0.99 | 0.98 | 0.019 | 0.98 |

TABLE. IV. THE RESULTS OF EVALUATION METHODS FOR k-NN

| | F1→1 | F1→0 | Weighted Avg → F1 | MSE | Accuracy |
|---|---|---|---|---|---|
| k-NN | 0.91 | 0.96 | 0.95 | 0.055 | 0.95 |

## IV. RESULTS INTERPRETATION AND EVALUATION

In this section, the results obtained from the previous sections for both Decision Tree (DT) and k-Nearest Neighbor k-NN algorithms will be interpreted and discussed in this section. In fact, more than one evaluation criteria have been used to evaluate the proposed model. Model evaluation is an integral part of the model development process, which helps to find the best model that represents the data and how well the chosen model will work in the future. The used evaluation measures are described in detail below:

- Accuracy: the most commonly used metric to judge a model. It can be measured according to the percentage of the recognized hand images per the total number of tested hand images. Accuracy can be calculated as following, where TP is the true positive instances, TN is the true negative instances, FP is the false positive instances, and FN is the false negative instances.

$$\text{Accuracy} = \frac{TP+TN}{TP+TN+FP+FN} \quad (1)$$

- F1 Score: also called F-measure, considers both the precision and the recall to compute the score. The F1 score is the harmonic mean of the precision and recall, where an F1 score reaches its best value at 1 (perfect precision and recall) and worst at 0.

$$\text{F1 Score} = \frac{2*\text{Precision}*\text{Recall}}{\text{Precision}+\text{Recall}} \quad (2)$$

- Mean Square Error (MSE): measures the average squared difference between the estimated values and true value. It is a risk function, corresponding to the expected value of the squared error loss, always nonnegative, values close to zero are better.

According to the following Table V., it has been found that DT has a higher accuracy with a value 0.98 than k-NN with a value 0.95, and higher F1 function with a value 0.98 comparing with k-NN with a value 0.95.

TABLE. V. COMPREHENSIVE RESULTS INTERPRETATION AND EVALUATION

| | F1→1 | F1→0 | Weighted Avg → F1 | MSE | Accuracy |
|---|---|---|---|---|---|
| Decision Tree | 0.97 | 0.99 | 0.98 | 0.019 | 0.98 |
| k-NN | 0.91 | 0.96 | 0.95 | 0.055 | 0.95 |

In term of Mean Square Error (MSE), DT has lower value 0.019 than k-NN with a value 0.055, which means that Decision Tree method has higher performance comparing with k-NN method to predict the state of the mosque's domes using weather features and outside temperatures.

The following Figure 4 shows the Confusion Matrix for k-NN and DT.

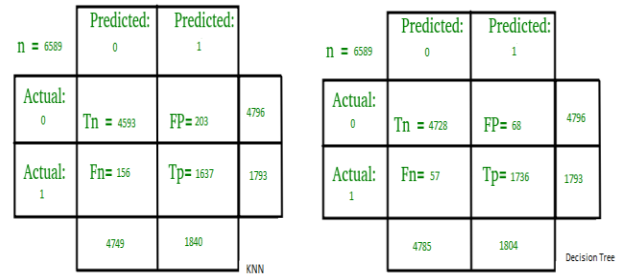

Fig. 4. Confusion Matrix for k-NN and DT.

After comparing all the results which have been achieved from Confusion Matrix as mentioned in Figure 4 for both models, it has been found that the DT algorithm was better method comparing with k-NN, so it will be used for building the smart domes model.

Given the results that we obtained previously, we can say that we have succeeded in building the smart domes system to the Prophet's Mosque in Saudi Arabia which can be expanded to other mosques all over the world, where we can now control the domes to make decisions automatically using machine learning algorithms considering some significant weather features and outside temperatures. In addition, the problems of ventilation, pollution, spread of germs and inappropriate smells in the Mosques have been solved successfully.

## V. CONCLUSION

In this paper, a system of smart domes has been proposed using weather features and outside temperatures. Machine learning algorithms such as k-NN and Decision Tree have been applied on weather features and outside temperature to predict the state of the mosque's domes (open or close). The results of this study are promising and will be helpful for all mosques to use our proposed model for controlling domes automatically.

Due to the difficulty of ventilating the mosques, decreasing the pollution, spread of germs and inappropriate smells in the Mosques, our proposed model will be very good to solve these problems and keep the mosques healthy for worshipers and visitors.

In the future work, some ideas will be applied to solve the problem of people crowding inside the mosque by determining the time for dome to be opened in minutes by specifying the number of worshipers inside the mosque using fuzzy based control, specially that fuzzy systems have been applied in many different industries [18]. In addition to applying some other machine and deep learning algorithms in the future to increase the performance of the model.

REFERENCES

[1] Deloitte and Dubai Islamic Economy Development Center. https://www.sasapost.com/largest-numbers-of-mosques/.

[2] General Presidency of al Haram Mosque and the Prophet's Mosque. https://www.gph.gov.sa/ar-sa/MasjidulNabawi/Pages/Building-and-expansion-Nabawi-Mosque.aspx.

[3] Nalanda B Dudde, Dr.S.S. Apte , "Arbitrary Decision Tree for Weather Prediction" , International Journal of Science and Research (IJSR) , Vol. 5 , Issue. 3,pp. 87-89, doi:10.21275/v5i3.nov161774. 2016.

[4] General Authority for Statistics 'kingdom of Saudi Arabia'. https://www.stats.gov.sa/ar/news/203.

[5] M.Kannan, S.Prabhakaran, P.Ramachandran, "Rainfall Forecasting Using Data Mining Technique", International Journal of Engineering and Technology Vol. 2, no. 6,pp. 397-401, 2010.

[6] A. Blasi, "Performance Increment of High School Students using ANN Model and SA Algorithm". Journal of Theoretical & Applied Information Technology, 95(11):2417-2425. 2017.

[7] B. M. Gupta and S. M. Dhawan, "Deep Learning Research: Scientometric Assessment of Global Publications Output during 2004 - 17," Emerging Science Journal, vol. 3, no. 1, p. 23, Feb. 2019. doi:10.28991/esj-2019-01165.

[8] S. Ghorbani, M. Barari, and M. Hosseini, "A Modern Method to Improve of Detecting and Categorizing Mechanism for Micro Seismic Events Data Using Boost Learning System," Aug. 2017. doi:10.20944/preprints201708.0072.v1.

[9] Saba, T., Rehman, A., & AlGhamdi, J. S. "Weather forecasting based on hybrid neural model". Applied Water Science, 7(7), 3869–3874, 2017. doi:10.1007/s13201-017-0538-0.

[10] Ch.Jyosthna Devi, B.Syam Prasad Reddy, K.Vagdhan Kumar,B.Musala Reddy,N.Raja Nayak , "ANN Approach for Weather Prediction using Back Propagation ", International Journal of Engineering Trends and Technology , Vol. 3 , Issue. 1,pp. 19-23, 2012.

[11] Nikhil Sethi, Dr.Kanwal Garg , "Exploiting Data Mining Technique for Rainfall Prediction", International Journal of Computer Science and Information Technologies, Vol. 5, No. 3,pp. 3982-3984, 2014.

[12] Saudi Arabia weather history data. Kaggle Website: https://www.kaggle.com/esraamadi/saudi-arabia-weather-history.

[13] Sayali D. Jadhav, H. P. Channe , "Comparative Study of K-NN, Naive Bayes and Decision Tree Classification Techniques ", International Journal of Science and Research (IJSR), Vol. 5 Issue. 1, pp. 1842-1845, 2016.

[14] S. Kumar, T. Roshni, and D. Himayoun, "A Comparison of Emotional Neural Network (ENN) and Artificial Neural Network (ANN) Approach for Rainfall-Runoff Modelling," Civil Engineering Journal, vol. 5, no. 10, pp. 2120–2130, Oct. 2019. doi:10.28991/cej-2019-03091398.

[15] J. Han, M. Kamber, and J. Pei, "Classification," Data Mining, pp. 327–391, 2012. doi:10.1016/b978-0-12-381479-1.00008-3.

[16] F. LI, Y. LI, and C. WANG, "Uncertain data decision tree classification algorithm," Journal of Computer Applications, vol. 29, no. 11, pp. 3092–3095, Dec. 2009.

[17] O. Kramer, "K-Nearest Neighbors," Intelligent Systems Reference Library, pp. 13–23, 2013.

[18] A. Blasi, "Scheduling food industry system using fuzzy logic". Journal of Theoretical & Applied Information Technology, 96(19): 6463-6473. 2018.